\documentclass[sigconf]{acmart}
\usepackage{algorithm}
\usepackage{algpseudocode}
\usepackage[capitalise, noabbrev]{cleveref}
\AtBeginDocument{%
  }

\setcopyright{acmlicensed}
\copyrightyear{2018}
\acmYear{2018}
\acmDOI{XXXXXXX.XXXXXXX}
\acmConference[Conference acronym 'XX]{Make sure to enter the correct
  conference title from your rights confirmation email}{June 03--05,
  2018}{Woodstock, NY}
\acmISBN{978-1-4503-XXXX-X/2018/06}

\acmSubmissionID{4705}

\begin{document}

\title{W2W: A Simulated Exploration of IMU Placement Across the Human Body for Designing Smarter Wearable}

\author{Lala Shakti Swarup Ray}
\affiliation{%
  \institution{DFKI, RPTU}
  \city{Kaiserslautern}
  \country{Germany}
  }
\email{lala_shakti_swarup.ray@dfki.de}
\orcid{0000-0002-7133-0205}

\author{Bo Zhou}
\affiliation{%
  \institution{DFKI, RPTU}
   \city{Kaiserslautern}
  \country{Germany}}
\email{bo.zhou@dfki.de}

\author{Paul Lukowicz}
\affiliation{%
  \institution{DFKI, RPTU}
  \city{Kaiserslautern}
  \country{Germany}}
\email{paul.lukowicz@dfki.de}

\renewcommand{\shortauthors}{Ray et al.}

\begin{abstract}
Inertial measurement units (IMUs) are central to wearable systems for activity recognition and pose estimation, but sensor placement remains largely guided by heuristics and convention. In this work, we introduce Where to Wear (W2W), a simulation-based framework for systematic exploration of IMU placement utility across the body. Using labeled motion capture data, W2W generates realistic synthetic IMU signals at 512 anatomically distributed surface patches, enabling high-resolution, task-specific evaluation of sensor performance.
We validate W2W’s reliability by comparing spatial performance rankings from synthetic data with real IMU recordings in two multimodal datasets, confirming strong agreement in activity-wise trends.
Further analysis reveals consistent spatial trends across activity types and uncovers overlooked high-utility regions that are rarely used in commercial systems. These findings challenge long-standing placement norms and highlight opportunities for more efficient, task-adaptive sensor configurations.
Overall, our results demonstrate that simulation with W2W can serve as a powerful design tool for optimizing sensor placement, enabling scalable, data-driven strategies that are impractical to obtain through physical experimentation alone. Upon acceptance, we plan to release W2W as an open-source web app.
\end{abstract}



\received{20 February 2007}
\received[revised]{12 March 2009}
\received[accepted]{5 June 2009}

\maketitle
\section{Introduction}

Wearable systems incorporating inertial measurement units (IMUs) have become central to a wide range of applications, including human activity recognition (HAR) \cite{nematallah2024adaptive, ray2024har}, motion tracking \cite{liu20243d}, rehabilitation \cite{felius2024exploring}, sports analytics \cite{biro2024ai}, and ubiquitous computing \cite{dhage2024machine}. 
Their portability and affordability have enabled broad adoption in both research and commercial devices.
However, the most optimal sensor placement on the body remains an underexplored questions, which directly influences the signal characteristics and representativeness of the activity motion, and most importantly, the performance of machine learning models depending on IMUs\cite{niswander2020optimization, sara2023effect}.
For example, the same activity can yield highly distinguishable signals when captured from a wrist-mounted sensor but may be difficult to classify from a sensor placed on the torso. 

Despite this importance, identifying ideal sensor locations is challenging; real-world experimentation is resource intensive, requiring custom hardware, calibration, and subject instrumentation. The cost scales linearly with placements and exponentially with sensor combinations, compounded by limitations in subject tolerance and data collection feasibility\cite{tan2019influence, anwary2018optimal}. 
Consequently, a comprehensive understanding of task-specific optimal placement, particularly for multi-sensor systems, is lacking.

To overcome these practical barriers, we introduce Where to Wear (W2W), a simulation-based framework enabling scalable, systematic computational evaluation of IMU sensor placements grounded in anatomical accuracy. 
Our approach leverages labeled motion capture (MoCap) data to synthesize realistic IMU signals across thousands of candidate body surface locations as given in \cref{fig:overview}. 
W2W synthesizes realistic IMU signals across over 512 candidate locations by leveraging labeled motion capture (MoCap) data and the SMPL body model \cite{loper2023smpl}. 
This approach generates spatial utility maps reflecting location informativeness for downstream tasks like activity classification.
W2W also taps into the large corpus of MoCap datasets which is magnitudes larger than the wearable HAR domain.
Crucially, we validate our simulation approach against real-world annotated MoCap + IMU datasets to ensure fidelity to real sensor data and demonstrate its ability to reveal task-dependent spatial utility patterns and inform efficient sensor subset selection.
Our approach provides insights that would be prohibitively difficult to obtain through physical experimentation alone.
W2W will be released publicly as an open-source resource for optimizing sensor placement in wearable systems.

In summary, our contributions are as follows:

\begin{figure*}[!t]
\centering
\includegraphics[width=0.8\linewidth]{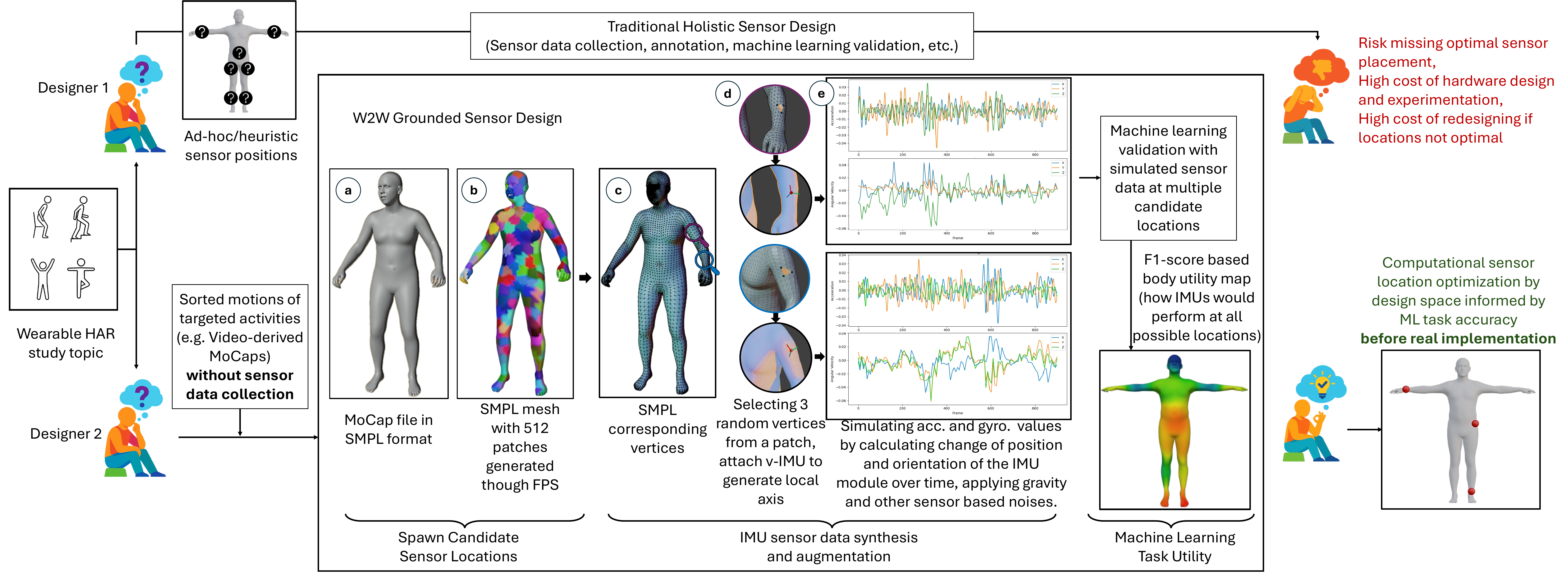}
\caption{W2W Sensor Design Workflow: It illustrates a simulation-driven pipeline for optimizing wearable sensor placement. Starting from representative rehabilitation activities (left), the workflow proceeds through:(a) Conversion of motion capture data to SMPL body mesh, (b) Generation of 512 anatomically distributed surface patches via farthest point sampling (FPS), (c) Patch-wise IMU modeling and sensor attachment simulation, (d) Per-location inertial data synthesis for multiple activities, and
(e) Evaluation of sensor signal quality across locations.
The rightmost panels contrast traditional sensor design uncertainty with W2W's data-driven sensor ranking and subset selection outputs.}
\label{fig:overview}
\end{figure*}

\begin{itemize}
  \item We present a novel simulation framework that generates realistic IMU data from MoCap input, densely evaluating multiple sensor locations over the full body surface.
  \item We validate the fidelity of W2W using multi-modal datasets MM-Fit \cite{stromback2020mm} and VIDIMU \cite{martinez2023vidimu} that include real IMU recordings from multiple body sites.
  \item We use W2W to identify task-specific sensor importance, efficient minimal sensor sets, and underutilized yet high-performing regions enabling placement analyses that are infeasible in traditional hardware-based workflows.
\end{itemize}

\section{Related  Work}
\subsubsection*{Sensor Placement Optimization Studies}
Prior studies have shown that IMU placement significantly influences the accuracy of activity recognition and motion estimation. 
Early work focused on a few sparsely predefined body sites such as the hip, wrist, or ankle, identified through empirical testing or user convenience \cite{cleland2013optimal, pannurat2017analysis, uhlenberg2024mount}. More recent efforts have explored optimal sensor configurations using data-driven methods. Notably, Tsukamoto et al. conducted the most fine-grained study to date, using a custom garment with 396 embedded IMUs to evaluate placement across the full body \cite{tsukamoto2023best}. While their findings provide valuable insights, their approach is based entirely on real sensor data, supports only a limited set of 14 activities, and requires expensive, highly specialized hardware making it difficult to extend to new activities or sensor positions. 
In contrast, Proposed W2W uses simulation to generate realistic IMU signals across any desired number of anatomically distributed locations, enabling scalable, low-cost, and task-adaptive placement analysis without the logistical overhead of physical trials.

\subsubsection*{Simulation for System Optimization}

Simulation has emerged as a powerful method for optimizing wearable systems, particularly in settings where real-world experimentation is constrained by cost, time, or sensor limitations. Frameworks such as IMUSim \cite{young2011imusim}, WIMUSim \cite{oishi2025wimusim}, CromoSim \cite{hao2022cromosim}, and DPSG \cite{zolfaghari2024sensor} generate synthetic IMU data from other widely available modalities. IMUSim, in particular, serves as a core simulation engine for data generation pipelines such as IMUTube \cite{kwon2020imutube} and IMUGPT \cite{leng2023generating, leng2024imugpt}, which synthesize dataset-specific IMU signals from video and text inputs to improve downstream HAR performance. Similarly, DiP \cite{huang2018deep}, IMUPoser \cite{mollyn2023imuposer}, and Loose Inertial Poser \cite{zuo2024loose} leverage SMPL-based virtual IMU simulation to train pose
estimation models at scale, demonstrating the versatility of synthetic inertial data for both classification and regression tasks. 
In parallel, recent work has extended simulation to tackle system-level optimization challenges, such as benchmarking motion capture techniques in loose-fitting garment scenarios \cite{ray2023selecting, ray2024comprehensive}. Extending this line of work, our W2W framework uses SMPL to systematically evaluate and rank IMU placements, offering data-driven guidance for sensor deployment in wearable systems.

\section{W2W Framework}
W2W is designed to generate realistic IMU signals at arbitrary locations on the human body using 3D motion data. It transforms pose sequencesinto synthetic accelerometer and gyroscope readings by modeling virtual sensors on the body surface. The framework consists of three main components: (1) synthetic IMU generation, (2) body surface mapping, and (3) noise and temporal modeling to replicate real-world sensor imperfections. This modular structure allows scalable and repeatable experimentation with sensor placement strategies, enabling insights that would be impossible to achieve through physical deployment.

\subsection{Synthetic IMU Generation}

To simulate IMU signals at arbitrary body surface locations, we leverage the SMPL model, which represents the human body as a deformable mesh with 6,890 vertices and a consistent topology across frames.

\textit{1. Sensor Initialization from Mesh Geometry:}
Given a 3D body pose (from motion capture or video), we convert it into SMPL format and extract the mesh at each timestep. To define a virtual IMU location, we identify the three nearest mesh vertices to the desired body location and form a triangle from them. This triangle serves as a local surface patch, and its properties are used to define the initial position and orientation of the sensor.
Let the three vertex positions at time \( t \) be \( \mathbf{v}_1(t), \mathbf{v}_2(t), \mathbf{v}_3(t) \). The sensor position is defined $\mathbf{p}(t) = \frac{1}{3} \left( \mathbf{v}_1(t) + \mathbf{v}_2(t) + \mathbf{v}_3(t) \right)$.
To compute orientation, we construct a local coordinate frame using the triangle's normal:
\begin{equation}
\mathbf{n}(t) = \frac{(\mathbf{v}_2(t) - \mathbf{v}_1(t)) \times (\mathbf{v}_3(t) - \mathbf{v}_1(t))}{\|(\mathbf{v}_2(t) - \mathbf{v}_1(t)) \times (\mathbf{v}_3(t) - \mathbf{v}_1(t))\|}
\end{equation}
This normal vector is used to define the sensor's "up" axis. The remaining axes can be constructed using an orthonormal basis from triangle edges.

\textit{2. Angular Velocity Estimation:}
To obtain angular velocity \( \boldsymbol{\omega}(t) \), we compute the relative rotation of the local frame between consecutive time steps. If \( \mathbf{R}(t) \) and \( \mathbf{R}(t - \Delta t) \) are the orientation matrices at two time steps, we compute:
\begin{equation}
\Delta \mathbf{R} = \mathbf{R}(t) \mathbf{R}(t - \Delta t)^{-1}
\end{equation}
Then, angular velocity is extracted using the matrix logarithm:
\begin{equation}
\boldsymbol{\omega}(t) = \frac{\log(\Delta \mathbf{R})^\vee}{\Delta t}
\end{equation}
where \( \log(\cdot)^\vee \) denotes the mapping from rotation matrix to axis-angle representation.

\textit{3. Linear Acceleration and Gravity Incorporation}
We estimate linear acceleration \( \mathbf{a}_{\text{local}}(t) \) by differentiating the position of the patch centroid twice:
\begin{equation}
\mathbf{a}_{\text{local}}(t) = \frac{\mathbf{v}(t + \Delta t) - 2\mathbf{v}(t) + \mathbf{v}(t - \Delta t)}{\Delta t^2}
\end{equation}

Since real IMUs measure acceleration inclusive of gravity, we incorporate the effect of gravity in the local sensor frame. Assuming global gravity \( \mathbf{g} = [0, -9.81, 0]^\top \), and using the local orientation matrix \( \mathbf{R}(t) \), we compute:
\begin{equation}
\mathbf{a}_{\text{imu}}(t) = \mathbf{a}_{\text{local}}(t) + \mathbf{R}(t)^{-1} \mathbf{g}
\end{equation}

This gives the final accelerometer reading \( \mathbf{a}_{\text{imu}}(t) \), simulating the data a real sensor would measure on a moving body.

\subsection{Body surface mapping strategy}
The SMPL model defines the human body as a triangulated mesh consisting of 6,890 vertices and approximately 13,776 faces. In theory, the number of possible unique IMU placement triangles from this vertex set is extremely large:
\begin{equation}
\text{Total combinations} = \binom{6890}{3} \approx 1.63 \times 10^{10}
\end{equation}
However, most of these combinations are either redundant, anatomically meaningless, or located too close together to provide distinct signal characteristics.

To address this, we adopt a principled surface sampling approach that selects a fixed number \( N \) of representative surface regions from the mesh. In our framework, we set \( N = 512 \), which strikes a balance between dense spatial coverage and simulation efficiency.

\subsubsection*{Surface Patch Selection via Farthest Point Sampling}

To ensure anatomically uniform sensor coverage, we apply \textit{farthest point sampling (FPS)} \cite{han2023quickfps} over the SMPL mesh using geodesic distance given in \cref{alg:fps}. FPS selects surface points that are maximally spaced apart, resulting in well-distributed patch centers across the body.

Once \( N \) patch centers are identified, we define a local surface region (e.g., a geodesic disk or triangle fan) around each. From this region, we select three vertices to form a local triangle, which defines the position and orientation of a virtual IMU.

\begin{algorithm}[H]
\caption{Farthest Point Sampling on SMPL Mesh (Geodesic Domain)}
\label{alg:fps}
\begin{algorithmic}[1]
\footnotesize
\State \textbf{Input:} Mesh \( M = (V, E) \), number of patches \( N \)
\State \textbf{Output:} Patch centers \( C = \{c_1, ..., c_N\} \)
\State Initialize \( C \gets \{ \text{random vertex } v_0 \in V \} \)
\While{ \( |C| < N \) }
    \For{each vertex \( v \in V \setminus C \)}
        \State Compute geodesic distance: 
        \[
        d(v) = \min_{c \in C} \text{GeodesicDist}(v, c)
        \]
    \EndFor
    \State Select \( v^* = \arg\max_{v} d(v) \)
    \State Add \( v^* \) to \( C \)
\EndWhile
\State \Return \( C \)
\end{algorithmic}
\end{algorithm}

Although we use \( N = 512 \) in our default configuration, the FPS-based sampling algorithm is fully parameterized and can be reused to generate any desired number of placement patches.

\subsection{Temporal Filtering and Noise Modeling}

While synthetic IMU signals generated from kinematic data are clean and precise, real-world sensors introduce various sources of noise and temporal artifacts. To bridge this gap and improve the realism of simulated signals, our framework incorporates configurable models of temporal filtering and sensor noise. We applied a low-pass Butterworth filter and resampled the signals to common IMU rates (e.g., 25–200 Hz) to emulate hardware bandwidth limitations and discrete sampling. We added zero-mean Gaussian noise to simulate sensor variability, and introduced slowly drifting biases as well as fixed axis misalignment to reflect long-term drift and calibration errors. Together, these components allow our simulator to generate IMU signals that closely mimic the temporal and noise characteristics of real inertial devices.

\section{Evaluation}
\subsection{Goals and Datasets}

We use three different datasets for two distinct evaluation goals: (1) validation of the realism of our synthetic data (MM‑Fit \cite{stromback2020mm}, VIDIMU \cite{martinez2023vidimu} ) and (2) Sensor placement utility analysis from the wearable system developers' point of view (Mixamo \cite{blackman2014rigging}):
\begin{itemize}
\item \textbf{MM‑Fit \cite{stromback2020mm}} with 11 structured exercise activities using 5 IMU-equipped consumer devices (e.g., phones, smartwatches, earbuds) across multiple users, with synchronized video ground truth.
\item \textbf{VIDIMU \cite{martinez2023vidimu}} that includes 13 dynamic daily activities from 16 participants wearing 10 XSens IMUs, with RGB video recordings in varied scenarios.
\item \textbf{Mixamo \cite{blackman2014rigging}} animation library contains over 3,000 skeletal motion sequences (without IMU data), from which we selected 589 diverse motions to evaluate sensor placement utility. These sequences were manually categorized into activity types including object manipulation, locomotion, asymmetric actions, full-body rotations, balance-intensive tasks, and communicative gestures.
\end{itemize}

\subsection{W2W Validation}
Before evaluating the utility of our simulator, we first validate its realism by comparing the spatial trends of synthetic IMU data against real-world measurements. Rather than focusing on raw signal similarity—which is highly sensitive to alignment and noise—we assess whether the simulator preserves the relative informativeness of different sensor placements for activity recognition.

\subsubsection*{Validation Approach}

We simulate virtual IMUs at the same anatomical locations as those used in each dataset, using the SMPL mesh and our synthetic signal pipeline. Each virtual sensor outputs temporally filtered, gravity-aware acceleration and gyroscope signals.

Rather than comparing raw signal shapes, we validate our simulator by checking whether it preserves the relative utility of sensor placements for activity recognition. For each dataset:

\begin{itemize}
  \item We train a classifier from Multi$^3$Net \cite{fortes2024enhancing} per sensor location, using fixed-length time windows to predict activity labels.
  \item We rank sensor locations by classification performance for each activity.
  \item We compare these rankings between real and synthetic data using Spearman correlation \cite{de2016comparing}.
\end{itemize}

This approach reveals whether synthetic data maintains the task-relevant spatial structure that governs sensor effectiveness.

\subsubsection*{Results}

\begin{figure}[!t]
\centering
\includegraphics[width=0.9\linewidth]{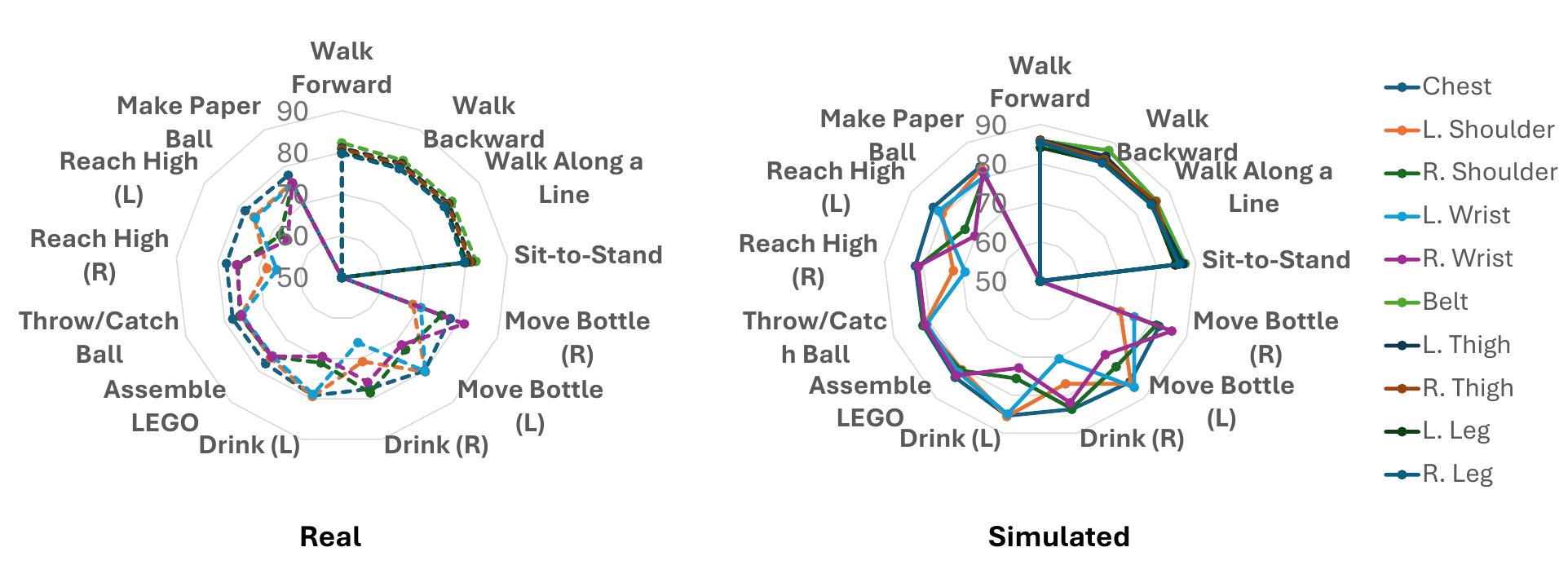}
\caption{Per-activity classification accuracy for each sensor location under real and synthetic data for VIDIMU dataset.}
\label{fig:validation1}
\end{figure}

\begin{figure}[!t]
\centering
\includegraphics[width=0.9\linewidth]{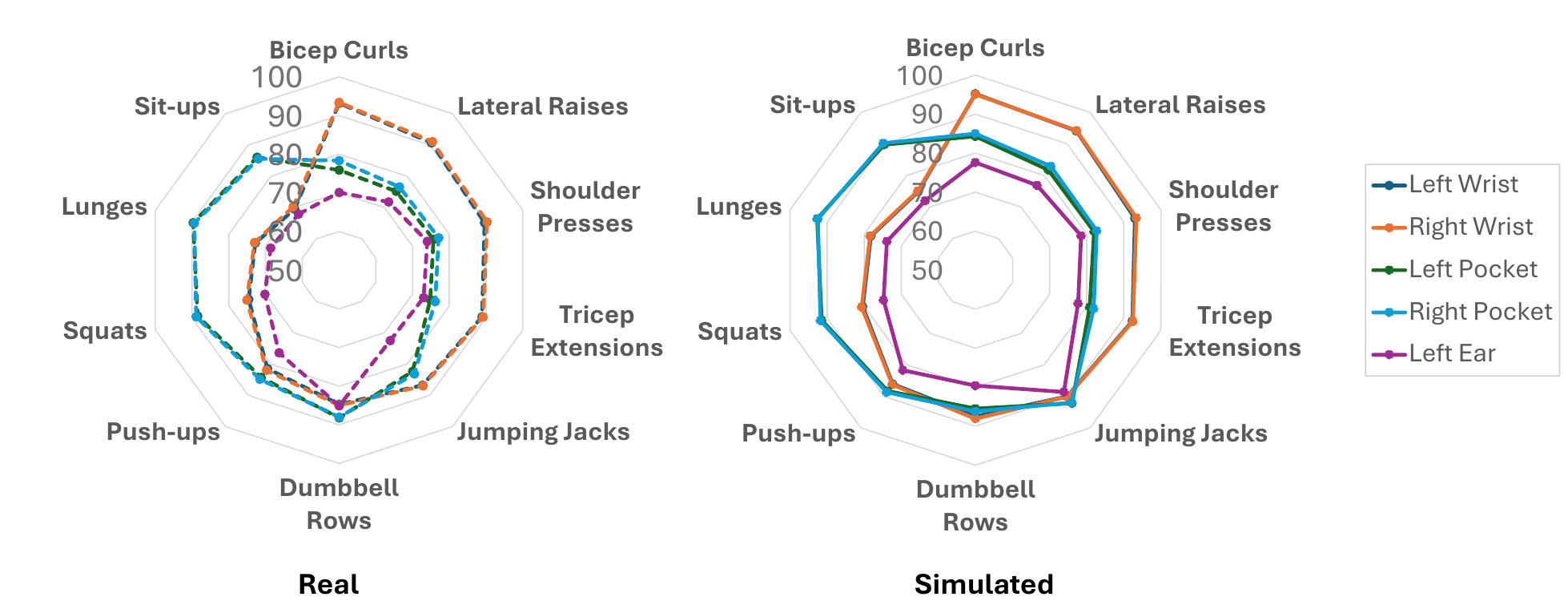}
\caption{Per-activity classification accuracy for each sensor location under real and synthetic data for MM-Fit dataset.}
\label{fig:validation2}
\end{figure}

Based on the observed results synthetic IMU data generally yields slightly higher F1 scores attributable to reduced noise and idealized dynamics but the key result is the preservation of spatial utility patterns. For each sensor, the relative performance across activities is consistent between real and synthetic conditions across datasets as visualized in \cref{fig:validation1} and \cref{fig:validation2}. Similarly, for each activity, the top-performing sensor regions are stable across modalities.
For instance, wrist-mounted sensors consistently lead in upper-body tasks like \textit{Bicep Curl} and \textit{Shoulder Press}, while thigh and hip-mounted sensors dominate locomotion tasks such as \textit{Squats} and \textit{Lunges}. These trends emerge consistently across datasets and conditions.

Despite differences in protocol, subject demographics, and sensor configurations, we observe strong agreement in utility patterns across datasets. Key joint regions (e.g., wrists, hips, thighs) consistently rank high in tasks involving those limbs, while uninvolved areas (e.g., the contralateral arm) rank lower.
These findings suggest that W2W not only generates structurally plausible inertial signals, but also preserves the underlying biomechanics that govern task-sensor relationships. Across datasets, we observe strong Spearman correlation ($\rho > 0.85$) in activity-wise sensor rankings between real and synthetic data.

This supports the use of simulation as a reliable proxy for physical prototyping in sensor placement studies—particularly when task-specific and body-region-specific trends are the primary focus.

\subsection{W2W Utility: Sensor Placement Analysis}

Following validation with real-world data, we used simulation to evaluate the spatial utility of IMU sensors across 512 anatomically distributed surface patches on the SMPL mesh. At each location, we trained a single-sensor classifier using simulated IMU data generated from diverse motion sequences from the Mixamo animation library \cite{blackman2014rigging}. We aggregated per-activity classification performance into fine-grained spatial utility maps, capturing the informativeness of each body region for specific motion types. We then analyzed the data for two key use cases (1) Task informed spatial performance and (2) Sensor position importance and redundancy analysis.

\begin{figure}[!t]
\centering
\includegraphics[width=0.9\linewidth]{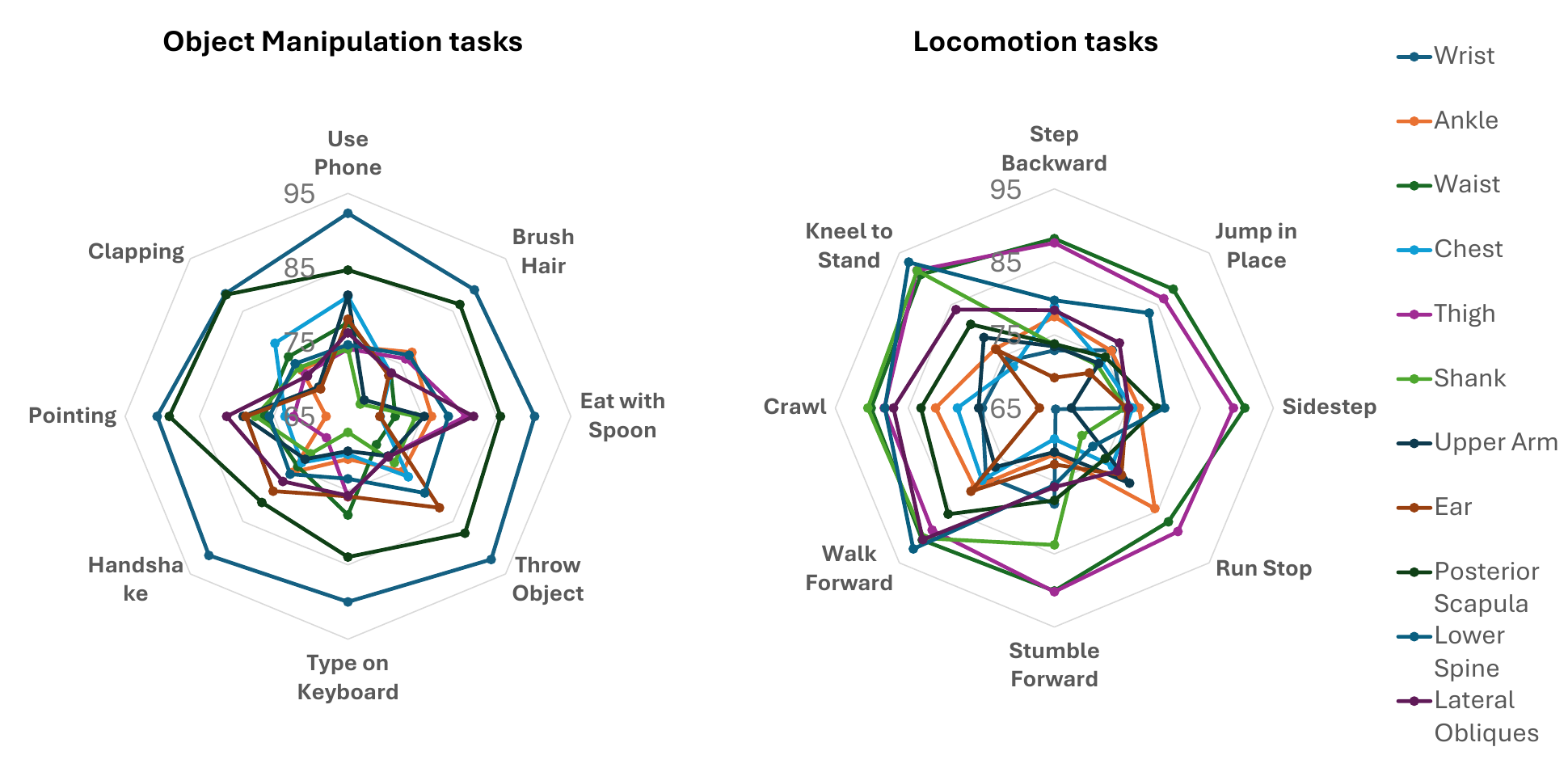}
\includegraphics[width=0.9\linewidth]{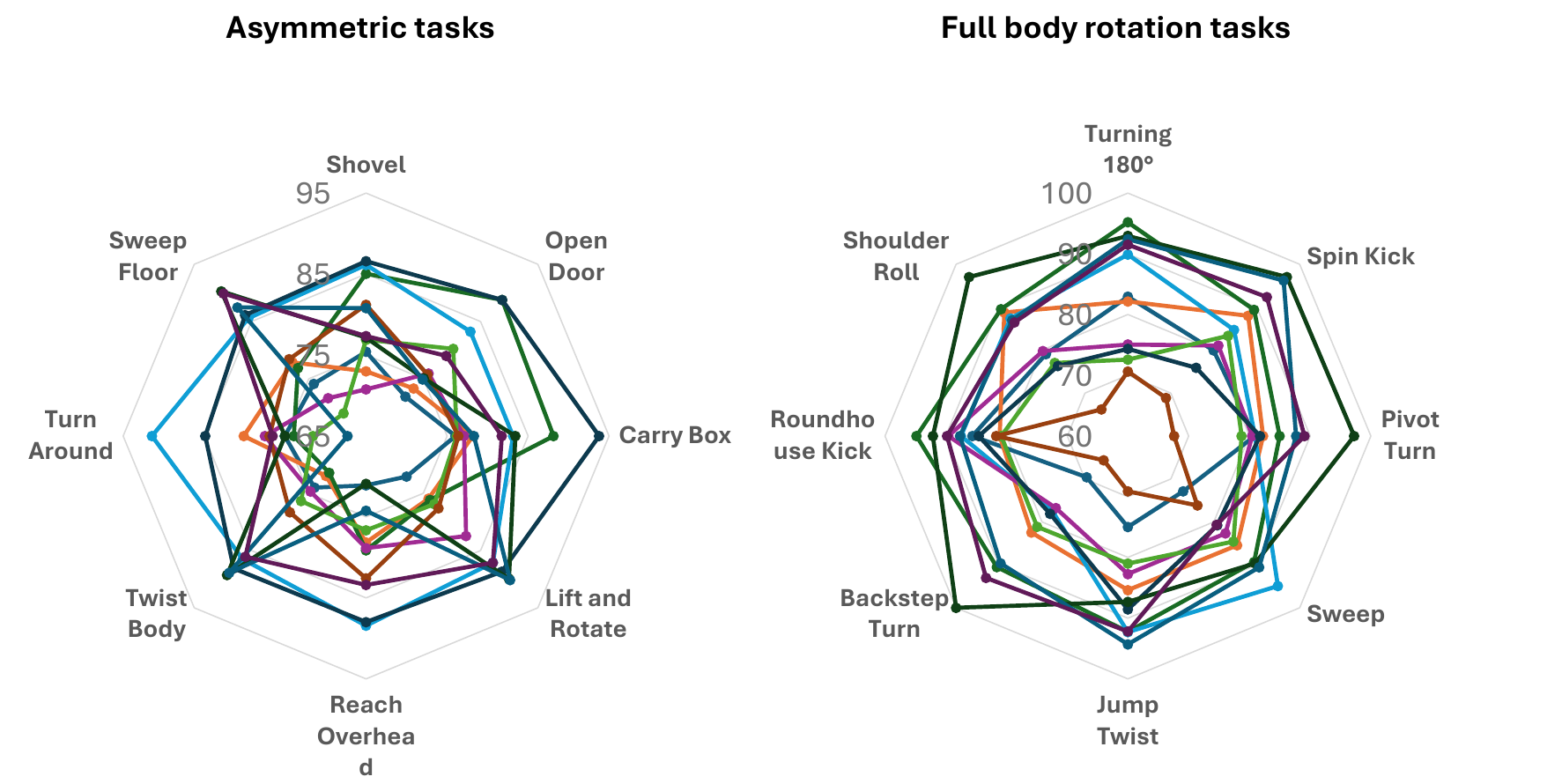}
\includegraphics[width=0.9\linewidth]{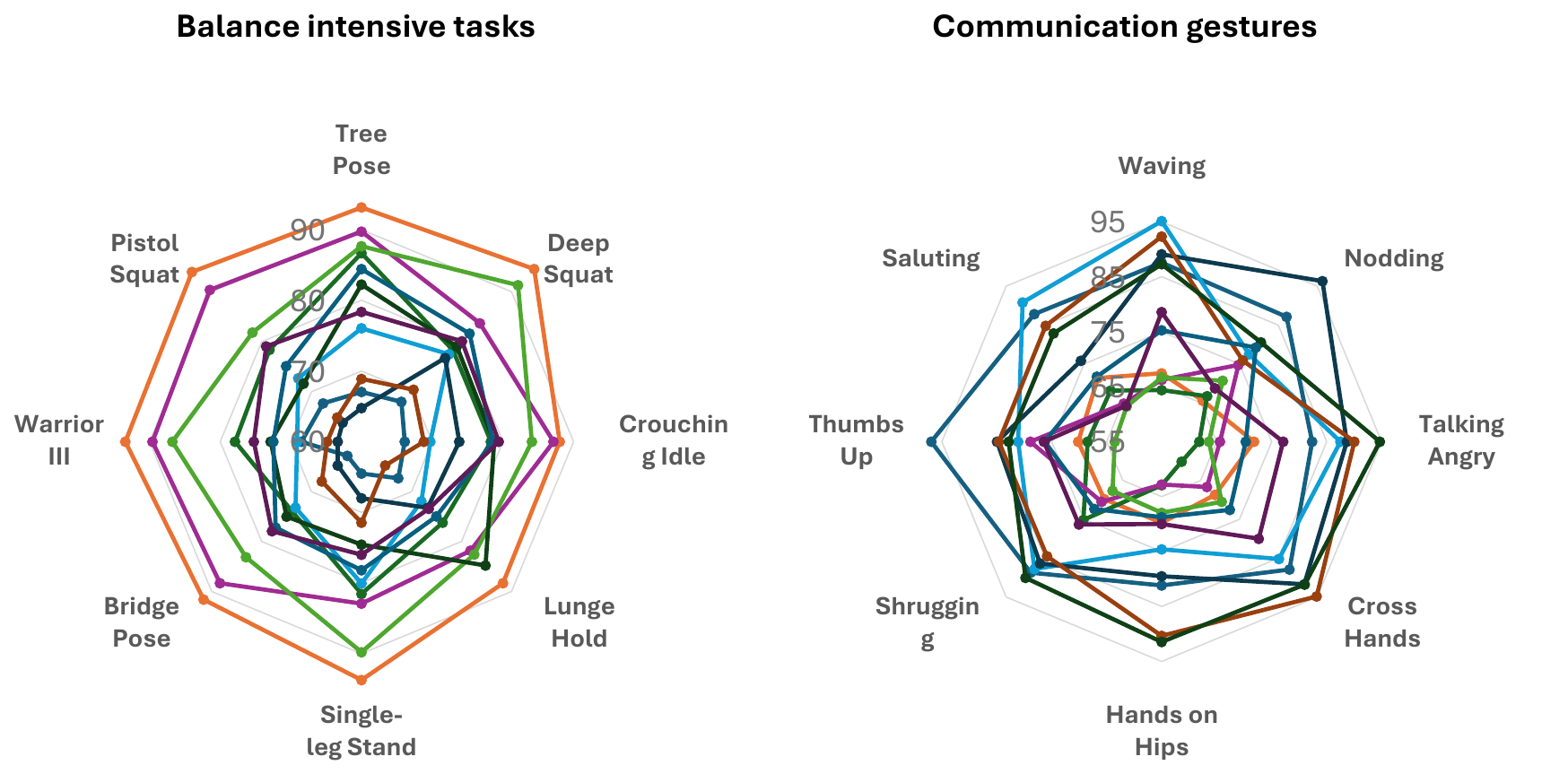}
\caption{F1-score results using W2W simulated IMU across 6 different activity groups (24 activities) from Mixamo and sensor placements (11/512), highlighting both common and underexplored configurations for different activity types.}
\label{fig:f1_scores_synthetic}
\end{figure}

\textit{1. Task-Informed Spatial Performance:} Simulations revealed interpretable trends in spatial utility, closely aligned with the kinematic demands of each task type, in particular we have demonstrated the position accuracy of 8 common and 3 unusual locations as given in \cref{fig:f1_scores_synthetic}. Based on the results \textit{forearm and hand regions} achieved the highest F1 scores in object-manipulation tasks such as \textit{Using Phone}, \textit{Throw Object}, and \textit{Eating with Spoon}, driven by high-frequency articulation at the distal joints. \textit{Thighs, shanks, and lower back} performed best in locomotion and posture-related tasks, including \textit{Crawling}, \textit{Walking}, and \textit{Kneeling}, where stable acceleration trajectories dominate.
\textit{Upper arms and hips} offered a favorable balance between signal utility and wearability in asymmetric tasks such as \textit{Load-Carrying}, \textit{Door-Opening}, and \textit{Shoveling}.
\textit{Pelvis, chest, and ankles} showed superior performance in full-body rotational tasks including \textit{Turning 180°}, \textit{Pivot Kicks}, and \textit{Spin Attacks}, effectively capturing coordinated yaw dynamics and stance transitions.
\textit{Lower back, thighs, and ankles} yielded strong performance in static and balance-intensive tasks, such as \textit{Tree Pose}, \textit{Deep Squat}, and \textit{Crouching Idle}, where postural stability is a key feature.
\textit{Forearms, shoulders, and head} performed best in expressive and communicative gestures, including \textit{Waving}, \textit{Nodding}, and \textit{Pointing}, which require both distal articulation and upper-body alignment.

Beyond the commonly used sites, W2W identified several high-performing yet underutilized regions, including the posterior scapula (upper back shoulder), lower spine (lower back), and lateral obliques (sides of the torso). The posterior scapula excelled in object manipulation and full-body rotation, while the lower spine and lateral obliques were especially effective for compound and rotational movements. Mechanically, these regions should stable signals with minimal motion-induced tissue artifacts. Ergonomically, they offer greater comfort and better concealment compared to sensors mounted on the limbs. These findings suggest they warrant consideration in future wearable designs.

\textit{Sensor Position Importance and Redundancy Analysis:}
Given a set of $T$ target activities, we generate $T$ heatmaps each mapping a location $l \in \{1, \dots, 512\}$ to an F1-score $F_{l}^{(t)} \in [0,1]$. From this, we support two key analyses:

(1) Optimal Single-Sensor Selection: To identify the best-performing single sensor location $l^*$ across all tasks, we compute the average F1-score for each location using $\bar{F}_l = \frac{1}{T} \sum_{t=1}^{T} F_l^{(t)}$.
The optimal location is defined as $l^* = \arg\max_l \bar{F}_l$.
This yields the body location that, on average, provides the most informative signal across the full activity set.

(2) Minimal Sensor Subset for Target Accuracy: We define a desired performance threshold $\tau \in [0,1]$ (e.g., 0.90). The objective is to identify the smallest subset $S \subseteq \{1,\dots,512\}$ such that the average maximum F1-score per activity exceeds $\tau$:

\begin{equation}
\text{Find } S^* = \arg\min_{S} |S| \quad \text{subject to } \frac{1}{T} \sum_{t=1}^{T} \max_{l \in S} F_l^{(t)} \geq \tau
\end{equation}

We approximate this combinatorial problem using a greedy algorithm given in \cref{alg:greedy_sensor_selection}. This greedy procedure iteratively selects a minimal sensor set that meets the target performance across all activities, supporting designs from single-sensor wearables to multi-sensor arrays for critical tasks. Such simulation-driven insights are challenging, if not infeasible, to derive through physical experimentation alone.

\begin{algorithm}
\vspace{-5pt}
\caption{Sensor Subset Selection for Target F1 Coverage}
\label{alg:greedy_sensor_selection}
\begin{algorithmic}[1]
\footnotesize
\State \textbf{Input:} F1-score matrix $F \in \mathbb{R}^{L \times T}$, threshold $\tau$
\State \textbf{Output:} Minimal sensor set $S$
\State $S \gets \emptyset$
\While{$\frac{1}{T} \sum_{t=1}^{T} \max_{l \in S} F_l^{(t)} < \tau$}
    \For{each $l \notin S$}
        \State Compute marginal gain:
        \[
        \Delta_l \gets \sum_{t=1}^{T} \max(F_l^{(t)}, \max_{j \in S} F_j^{(t)}) - \sum_{t=1}^{T} \max_{j \in S} F_j^{(t)}
        \]
    \EndFor
    \State Add $l = \arg\max \Delta_l$ to $S$
\EndWhile
\State \Return $S$
\end{algorithmic}
\end{algorithm}

\begin{figure}[!t]
\centering
\includegraphics[width=0.8\linewidth]{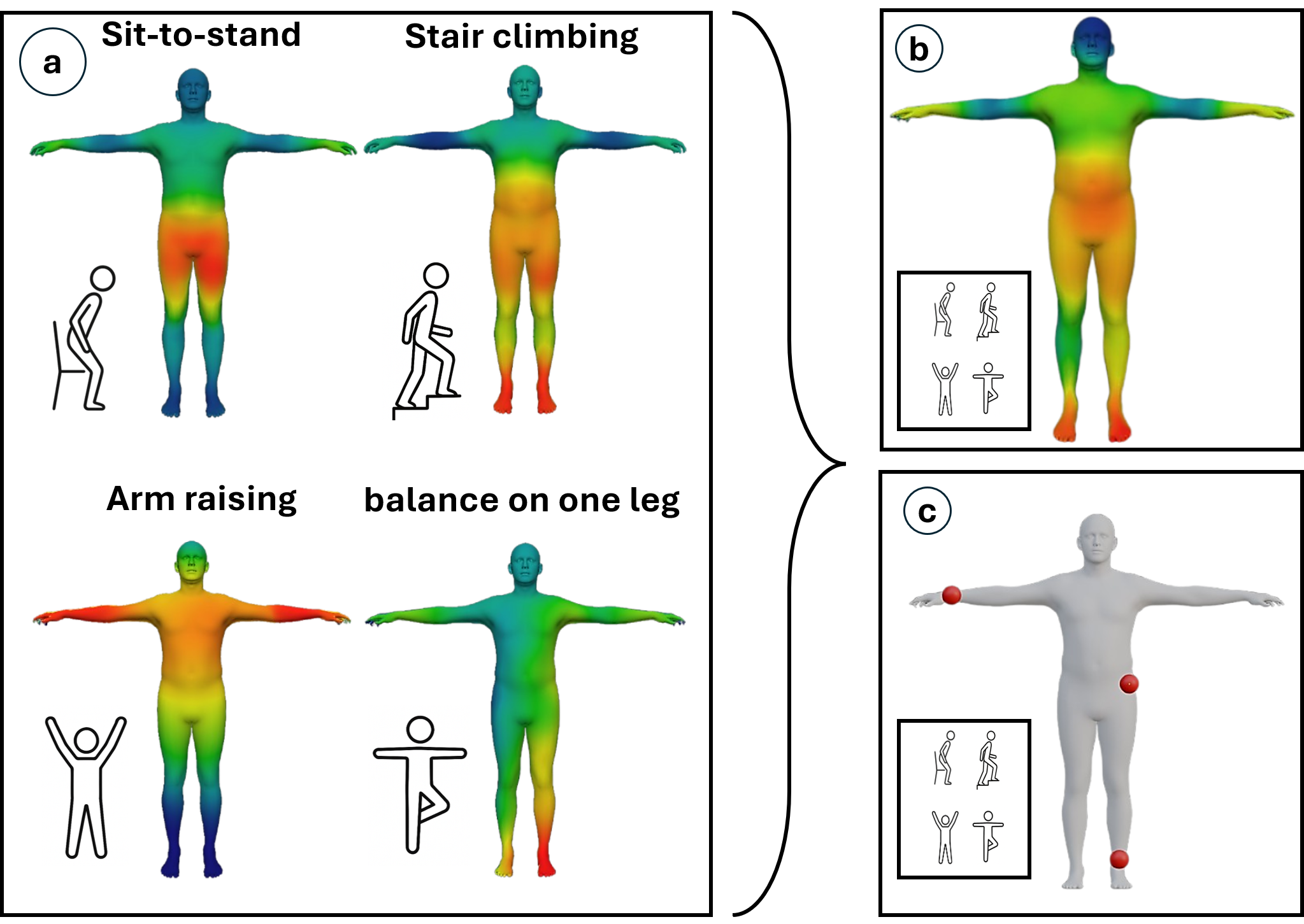}
\caption{Use case example of sensor position importance and redundancy analysis: (a) Per-activity heatmaps of classification performance across all sensor placements for the dataset with 4 activities.(b) Estimated sensor performance for the whole dataset across all sensor placements. (c) One possible solution minimum sensor combination to achieve at-least 90 \% F1 score for all activities in the dataset.}
\vspace{-0.4cm}
\label{fig:utilitymap}
\end{figure}

\textit{\textbf{Use Case Example} (Wearables for Rehab Design) :} 
To demonstrate W2W’s practical usage, we consider a rehabilitation scenario requiring recognition of four common at-home therapy activities: 
(1) Sit-to-stand, 
(2) Stair climbing, 
(3) Arm raises with a resistance band, and 
(4) One-leg standing balance. 
Each targets distinct muscle groups and movement patterns, necessitating accurate tracking of recovery monitoring. To design a wearable that can robustly recognize all four activities, we apply W2W in two stages:

\textit{(1) Sensor Placement Exploration:}  
We begin by analyzing spatial utility maps generated by simulating IMU data across 512 candidate locations shown in \Cref{fig:utilitymap}. Data augmentation provides more IMU data samples than motion files for a better ML training and validation process. Results indicate high F1-scores in the lower back and thighs for sit-to-stand and stair climbing, the forearms and shoulders for arm raises, and the ankles and hips for balance assessment. 
These trends are consistent with our earlier findings. Notably, the right lower back consistently achieves the highest average F1-score across these tasks, making it the most informative single location overall.

\textit{(2) Minimal Sensor Subset Selection for Target Performance:}  
Assuming a design goal of achieving at least a 90\% average F1-score across all four activities, we apply our greedy subset selection algorithm to minimize the number of required sensors. The algorithm selects a compact three-sensor configuration: right lower back, left forearm, and right ankle as provided in \cref{fig:utilitymap}.
This simulation-driven design strategy avoids the need for exhaustive physical prototyping, supports rapid iteration, and offers a principled basis for sensor placement in domain-specific wearable systems.

\section{Discussion and Limitation}

\subsubsection*{Why Simulation? Addressing the Limits of Real-World Data}

Despite the widespread use of IMUs, real-world sensor placement studies remain constrained by physical limitations: sensor mounting logistics, participant fatigue, limited body coverage, and restricted sensor counts. Our simulation-based framework fundamentally shifts this paradigm enabling high-resolution, scalable, and reproducible exploration of sensor configurations, free from the constraints of physical experimentation.
Simulation affords controlled variability, repeatability, and efficient large-scale experimentation. It allows researchers to explore questions that are impractical in hardware-based studies, such as: What is the minimal sensor layout required for a given task? How do utility trends shift across activity types or body morphologies? What is the marginal benefit of adding a new sensor? These capabilities support a more nuanced, task-driven understanding of sensor placement.

\subsubsection*{Rethinking Conventional Placement Norms}

Common IMU configurations such as placements on the wrists, ankles, or ear are often inherited from consumer wearable conventions rather than selected through empirical optimization. Our findings show that while these locations may offer convenience, they are not universally optimal.
Our simulations reveal underutilized regions such as the upper scapula, lower spine, and lateral torso that outperform traditional placements in certain task categories. These locations provide robust signals with minimal interference and can offer better comfort, concealment, or biomechanical stability. By systematically evaluating utility across the full body, simulation enables data-driven challenges to entrenched design norms.

\subsubsection*{Simulation as a Sensor Design Tool}

W2W supports rapid, reproducible comparison of hundreds of sensor placements across diverse motion types. It enables task-specific importance ranking, robustness analysis under varying kinematic conditions (e.g., speed, asymmetry), and efficient sensor subset selection and more provided in \cref{fig:table}. These capabilities make it particularly valuable during the early stages of wearable system design, where hardware iteration is costly or infeasible.
By decoupling exploration from hardware constraints, simulation accelerates the prototyping pipeline. Designers can identify promising configurations, test edge cases, and evaluate trade-offs before committing to physical deployments. This is especially critical in domains such as rehabilitation, industrial ergonomics, and sports, where task-specific sensor strategies are essential and user populations may be difficult to recruit.

\begin{table}[!t]
\centering
\caption{Major difference between contemporary wearable design vs W2W grounded sensor design.}
\includegraphics[width=0.8\linewidth]{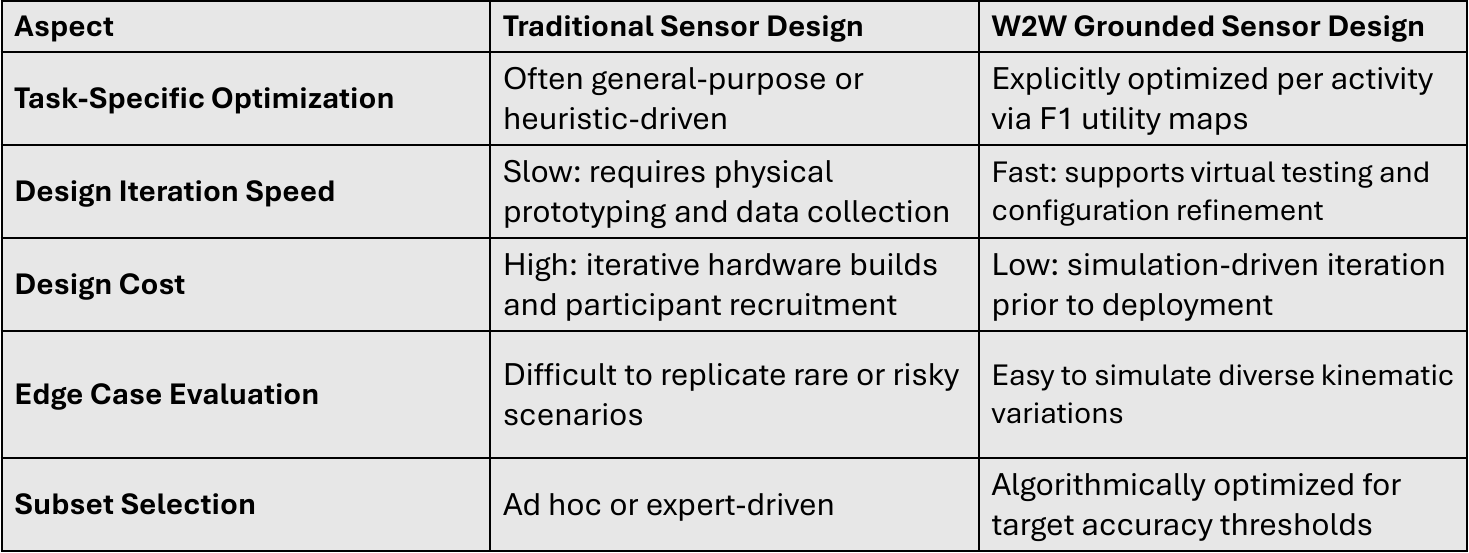}
\label{fig:table}
\vspace{-0.5cm}
\end{table}

\subsubsection*{Limitations and Future Directions}

Our simulator models body kinematics using motion capture-derived SMPL mesh sequences and assumes idealized rigid-body coupling for virtual IMUs. This abstraction enables scalable, consistent evaluation but does not account for real-world factors such as sensor slippage or environmental noise. While our current work focuses on HAR, W2W can support a broader range of downstream tasks like pose estimation, gait analysis, and motion quality assessment applications where accurate sensor placement can directly impact system performance. Ultimately, we envision a hybrid design workflow, where simulation augments but does not replace empirical testing. As fidelity improves, simulation can become a core component of end-to-end wearable design pipelines: informing placement decisions, guiding personalization strategies, and accelerating deployment across varied users and tasks.

\section{Conclusion}

W2W provides a computational simulation-based framework to systematically find optimal IMU sensor placements on the body to have maximum ML performance. By leveraging motion capture-driven synthesis, our approach enables dense, anatomically informed sampling of candidate locations far beyond the limits of real-world experimentation. We validated the fidelity and utility of W2W with two real-world multimodal datasets, and demonstrated that it accurately approximates location importance for specific HAR tasks. 

As demonstrated in the Use Case Example, wearable system practitioners can use W2W to evaluate the most optimal position so place IMU sensors for their specific tasks before real system implementation, by only providing assorted (possibly video-derived) motions.
Our approach not only cut cost and risk for evaluating new wearable study ideas, but also provides a systematic view of sensor utility for all possible locations, which could even reveal unconventional positions.

\begin{acks}
The research reported in this paper was supported by the BMBF in the project CrossAct (01IW21003).
\end{acks}

\bibliographystyle{ACM-Reference-Format}
\bibliography{sample-base}


\begin{thebibliography}{33}


\ifx \showCODEN    \undefined \def \showCODEN     #1{\unskip}     \fi
\ifx \showISBNx    \undefined \def \showISBNx     #1{\unskip}     \fi
\ifx \showISBNxiii \undefined \def \showISBNxiii  #1{\unskip}     \fi
\ifx \showISSN     \undefined \def \showISSN      #1{\unskip}     \fi
\ifx \showLCCN     \undefined \def \showLCCN      #1{\unskip}     \fi
\ifx \shownote     \undefined \def \shownote      #1{#1}          \fi
\ifx \showarticletitle \undefined \def \showarticletitle #1{#1}   \fi
\ifx \showURL      \undefined \def \showURL       {\relax}        \fi
\providecommand\bibfield[2]{#2}
\providecommand\bibinfo[2]{#2}
\providecommand\natexlab[1]{#1}
\providecommand\showeprint[2][]{arXiv:#2}

\bibitem[Anwary et~al\mbox{.}(2018)]%
        {anwary2018optimal}
\bibfield{author}{\bibinfo{person}{Arif~Reza Anwary}, \bibinfo{person}{Hongnian Yu}, {and} \bibinfo{person}{Michael Vassallo}.} \bibinfo{year}{2018}\natexlab{}.
\newblock \showarticletitle{Optimal foot location for placing wearable IMU sensors and automatic feature extraction for gait analysis}.
\newblock \bibinfo{journal}{\emph{IEEE Sensors Journal}} \bibinfo{volume}{18}, \bibinfo{number}{6} (\bibinfo{year}{2018}), \bibinfo{pages}{2555--2567}.
\newblock


\bibitem[Bir{\'o} et~al\mbox{.}(2024)]%
        {biro2024ai}
\bibfield{author}{\bibinfo{person}{Attila Bir{\'o}}, \bibinfo{person}{Antonio~Ignacio Cuesta-Vargas}, {and} \bibinfo{person}{L{\'a}szl{\'o} Szil{\'a}gyi}.} \bibinfo{year}{2024}\natexlab{}.
\newblock \showarticletitle{AI-assisted fatigue and stamina control for performance sports on IMU-generated multivariate times series datasets}.
\newblock \bibinfo{journal}{\emph{Sensors}} \bibinfo{volume}{24}, \bibinfo{number}{1} (\bibinfo{year}{2024}), \bibinfo{pages}{132}.
\newblock


\bibitem[Blackman and Blackman(2014)]%
        {blackman2014rigging}
\bibfield{author}{\bibinfo{person}{Sue Blackman} {and} \bibinfo{person}{Sue Blackman}.} \bibinfo{year}{2014}\natexlab{}.
\newblock \showarticletitle{Rigging with mixamo}.
\newblock \bibinfo{journal}{\emph{Unity for Absolute Beginners}} (\bibinfo{year}{2014}), \bibinfo{pages}{565--573}.
\newblock


\bibitem[Cleland et~al\mbox{.}(2013)]%
        {cleland2013optimal}
\bibfield{author}{\bibinfo{person}{Ian Cleland}, \bibinfo{person}{Basel Kikhia}, \bibinfo{person}{Chris Nugent}, \bibinfo{person}{Andrey Boytsov}, \bibinfo{person}{Josef Hallberg}, \bibinfo{person}{K{\aa}re Synnes}, \bibinfo{person}{Sally McClean}, {and} \bibinfo{person}{Dewar Finlay}.} \bibinfo{year}{2013}\natexlab{}.
\newblock \showarticletitle{Optimal placement of accelerometers for the detection of everyday activities}.
\newblock \bibinfo{journal}{\emph{Sensors}} \bibinfo{volume}{13}, \bibinfo{number}{7} (\bibinfo{year}{2013}), \bibinfo{pages}{9183--9200}.
\newblock


\bibitem[De~Winter et~al\mbox{.}(2016)]%
        {de2016comparing}
\bibfield{author}{\bibinfo{person}{Joost~CF De~Winter}, \bibinfo{person}{Samuel~D Gosling}, {and} \bibinfo{person}{Jeff Potter}.} \bibinfo{year}{2016}\natexlab{}.
\newblock \showarticletitle{Comparing the Pearson and Spearman correlation coefficients across distributions and sample sizes: A tutorial using simulations and empirical data.}
\newblock \bibinfo{journal}{\emph{Psychological methods}} \bibinfo{volume}{21}, \bibinfo{number}{3} (\bibinfo{year}{2016}), \bibinfo{pages}{273}.
\newblock


\bibitem[Dhage and Shankpale(2024)]%
        {dhage2024machine}
\bibfield{author}{\bibinfo{person}{Pragati Dhage} {and} \bibinfo{person}{Shubhangi Shankpale}.} \bibinfo{year}{2024}\natexlab{}.
\newblock \showarticletitle{Machine Learning-Driven Analysis of Smartphone IMU Data for Position and Step Counting}. In \bibinfo{booktitle}{\emph{2024 12th International Conference on Intelligent Systems and Embedded Design (ISED)}}. IEEE, \bibinfo{pages}{1--6}.
\newblock


\bibitem[Felius et~al\mbox{.}(2024)]%
        {felius2024exploring}
\bibfield{author}{\bibinfo{person}{Richard Felius}, \bibinfo{person}{Michiel Punt}, \bibinfo{person}{Marieke Geerars}, \bibinfo{person}{Natasja Wouda}, \bibinfo{person}{Rins Rutgers}, \bibinfo{person}{Sjoerd Bruijn}, \bibinfo{person}{Sina David}, {and} \bibinfo{person}{Jaap van Die{\"e}n}.} \bibinfo{year}{2024}\natexlab{}.
\newblock \showarticletitle{Exploring unsupervised feature extraction of IMU-based gait data in stroke rehabilitation using a variational autoencoder}.
\newblock \bibinfo{journal}{\emph{Plos one}} \bibinfo{volume}{19}, \bibinfo{number}{10} (\bibinfo{year}{2024}), \bibinfo{pages}{e0304558}.
\newblock


\bibitem[Fortes~Rey et~al\mbox{.}(2024)]%
        {fortes2024enhancing}
\bibfield{author}{\bibinfo{person}{Vitor Fortes~Rey}, \bibinfo{person}{Lala Shakti~Swarup Ray}, \bibinfo{person}{Qingxin Xia}, \bibinfo{person}{Kaishun Wu}, {and} \bibinfo{person}{Paul Lukowicz}.} \bibinfo{year}{2024}\natexlab{}.
\newblock \showarticletitle{Enhancing Inertial Hand based HAR through Joint Representation of Language, Pose and Synthetic IMUs}. In \bibinfo{booktitle}{\emph{Proceedings of the 2024 ACM International Symposium on Wearable Computers}}. \bibinfo{pages}{25--31}.
\newblock


\bibitem[Han et~al\mbox{.}(2023)]%
        {han2023quickfps}
\bibfield{author}{\bibinfo{person}{Meng Han}, \bibinfo{person}{Liang Wang}, \bibinfo{person}{Limin Xiao}, \bibinfo{person}{Hao Zhang}, \bibinfo{person}{Chenhao Zhang}, \bibinfo{person}{Xiangrong Xu}, {and} \bibinfo{person}{Jianfeng Zhu}.} \bibinfo{year}{2023}\natexlab{}.
\newblock \showarticletitle{QuickFPS: Architecture and algorithm co-design for farthest point sampling in large-scale point clouds}.
\newblock \bibinfo{journal}{\emph{IEEE Transactions on Computer-Aided Design of Integrated Circuits and Systems}} \bibinfo{volume}{42}, \bibinfo{number}{11} (\bibinfo{year}{2023}), \bibinfo{pages}{4011--4024}.
\newblock


\bibitem[Hao et~al\mbox{.}(2022)]%
        {hao2022cromosim}
\bibfield{author}{\bibinfo{person}{Yujiao Hao}, \bibinfo{person}{Xijian Lou}, \bibinfo{person}{Boyu Wang}, {and} \bibinfo{person}{Rong Zheng}.} \bibinfo{year}{2022}\natexlab{}.
\newblock \showarticletitle{Cromosim: A deep learning-based cross-modality inertial measurement simulator}.
\newblock \bibinfo{journal}{\emph{IEEE Transactions on Mobile Computing}} \bibinfo{volume}{23}, \bibinfo{number}{1} (\bibinfo{year}{2022}), \bibinfo{pages}{302--312}.
\newblock


\bibitem[Huang et~al\mbox{.}(2018)]%
        {huang2018deep}
\bibfield{author}{\bibinfo{person}{Yinghao Huang}, \bibinfo{person}{Manuel Kaufmann}, \bibinfo{person}{Emre Aksan}, \bibinfo{person}{Michael~J Black}, \bibinfo{person}{Otmar Hilliges}, {and} \bibinfo{person}{Gerard Pons-Moll}.} \bibinfo{year}{2018}\natexlab{}.
\newblock \showarticletitle{Deep inertial poser: Learning to reconstruct human pose from sparse inertial measurements in real time}.
\newblock \bibinfo{journal}{\emph{ACM Transactions on Graphics (TOG)}} \bibinfo{volume}{37}, \bibinfo{number}{6} (\bibinfo{year}{2018}), \bibinfo{pages}{1--15}.
\newblock


\bibitem[Kwon et~al\mbox{.}(2020)]%
        {kwon2020imutube}
\bibfield{author}{\bibinfo{person}{Hyeokhyen Kwon}, \bibinfo{person}{Catherine Tong}, \bibinfo{person}{Harish Haresamudram}, \bibinfo{person}{Yan Gao}, \bibinfo{person}{Gregory~D Abowd}, \bibinfo{person}{Nicholas~D Lane}, {and} \bibinfo{person}{Thomas Ploetz}.} \bibinfo{year}{2020}\natexlab{}.
\newblock \showarticletitle{Imutube: Automatic extraction of virtual on-body accelerometry from video for human activity recognition}.
\newblock \bibinfo{journal}{\emph{Proceedings of the ACM on Interactive, Mobile, Wearable and Ubiquitous Technologies}} \bibinfo{volume}{4}, \bibinfo{number}{3} (\bibinfo{year}{2020}), \bibinfo{pages}{1--29}.
\newblock


\bibitem[Leng et~al\mbox{.}(2024)]%
        {leng2024imugpt}
\bibfield{author}{\bibinfo{person}{Zikang Leng}, \bibinfo{person}{Amitrajit Bhattacharjee}, \bibinfo{person}{Hrudhai Rajasekhar}, \bibinfo{person}{Lizhe Zhang}, \bibinfo{person}{Elizabeth Bruda}, \bibinfo{person}{Hyeokhyen Kwon}, {and} \bibinfo{person}{Thomas Pl{\"o}tz}.} \bibinfo{year}{2024}\natexlab{}.
\newblock \showarticletitle{Imugpt 2.0: Language-based cross modality transfer for sensor-based human activity recognition}.
\newblock \bibinfo{journal}{\emph{Proceedings of the ACM on Interactive, Mobile, Wearable and Ubiquitous Technologies}} \bibinfo{volume}{8}, \bibinfo{number}{3} (\bibinfo{year}{2024}), \bibinfo{pages}{1--32}.
\newblock


\bibitem[Leng et~al\mbox{.}(2023)]%
        {leng2023generating}
\bibfield{author}{\bibinfo{person}{Zikang Leng}, \bibinfo{person}{Hyeokhyen Kwon}, {and} \bibinfo{person}{Thomas Pl{\"o}tz}.} \bibinfo{year}{2023}\natexlab{}.
\newblock \showarticletitle{Generating virtual on-body accelerometer data from virtual textual descriptions for human activity recognition}. In \bibinfo{booktitle}{\emph{Proceedings of the 2023 ACM International Symposium on Wearable Computers}}. \bibinfo{pages}{39--43}.
\newblock


\bibitem[Liu et~al\mbox{.}(2024)]%
        {liu20243d}
\bibfield{author}{\bibinfo{person}{Liujun Liu}, \bibinfo{person}{Jiewen Yang}, \bibinfo{person}{Ye Lin}, \bibinfo{person}{Peixuan Zhang}, {and} \bibinfo{person}{Lihua Zhang}.} \bibinfo{year}{2024}\natexlab{}.
\newblock \showarticletitle{3D human pose estimation with single image and inertial measurement unit (IMU) sequence}.
\newblock \bibinfo{journal}{\emph{Pattern Recognition}}  \bibinfo{volume}{149} (\bibinfo{year}{2024}), \bibinfo{pages}{110175}.
\newblock


\bibitem[Loper et~al\mbox{.}(2023)]%
        {loper2023smpl}
\bibfield{author}{\bibinfo{person}{Matthew Loper}, \bibinfo{person}{Naureen Mahmood}, \bibinfo{person}{Javier Romero}, \bibinfo{person}{Gerard Pons-Moll}, {and} \bibinfo{person}{Michael~J Black}.} \bibinfo{year}{2023}\natexlab{}.
\newblock \showarticletitle{SMPL: A skinned multi-person linear model}.
\newblock In \bibinfo{booktitle}{\emph{Seminal Graphics Papers: Pushing the Boundaries, Volume 2}}. \bibinfo{pages}{851--866}.
\newblock


\bibitem[Mart{\'\i}nez-Zarzuela et~al\mbox{.}(2023)]%
        {martinez2023vidimu}
\bibfield{author}{\bibinfo{person}{Mario Mart{\'\i}nez-Zarzuela}, \bibinfo{person}{Javier Gonz{\'a}lez-Alonso}, \bibinfo{person}{M{\'\i}riam Ant{\'o}n-Rodr{\'\i}guez}, \bibinfo{person}{Francisco~J D{\'\i}az-Pernas}, \bibinfo{person}{Henning M{\"u}ller}, {and} \bibinfo{person}{Cristina Sim{\'o}n-Mart{\'\i}nez}.} \bibinfo{year}{2023}\natexlab{}.
\newblock \showarticletitle{VIDIMU. Multimodal video and IMU kinematic dataset on daily life activities using affordable devices}.
\newblock \bibinfo{journal}{\emph{VIDIMU. Multimodal video and IMU kinematic dataset on daily life activities using affordable devices}} (\bibinfo{year}{2023}).
\newblock


\bibitem[Mollyn et~al\mbox{.}(2023)]%
        {mollyn2023imuposer}
\bibfield{author}{\bibinfo{person}{Vimal Mollyn}, \bibinfo{person}{Riku Arakawa}, \bibinfo{person}{Mayank Goel}, \bibinfo{person}{Chris Harrison}, {and} \bibinfo{person}{Karan Ahuja}.} \bibinfo{year}{2023}\natexlab{}.
\newblock \showarticletitle{Imuposer: Full-body pose estimation using imus in phones, watches, and earbuds}. In \bibinfo{booktitle}{\emph{Proceedings of the 2023 CHI Conference on Human Factors in Computing Systems}}. \bibinfo{pages}{1--12}.
\newblock


\bibitem[Nematallah and Rajan(2024)]%
        {nematallah2024adaptive}
\bibfield{author}{\bibinfo{person}{Heba Nematallah} {and} \bibinfo{person}{Sreeraman Rajan}.} \bibinfo{year}{2024}\natexlab{}.
\newblock \showarticletitle{Adaptive Hierarchical Classification for Human Activity Recognition Using Inertial Measurement Unit (IMU) Time-Series Data}.
\newblock \bibinfo{journal}{\emph{IEEE Access}} (\bibinfo{year}{2024}).
\newblock


\bibitem[Niswander et~al\mbox{.}(2020)]%
        {niswander2020optimization}
\bibfield{author}{\bibinfo{person}{Wesley Niswander}, \bibinfo{person}{Wei Wang}, {and} \bibinfo{person}{Kimberly Kontson}.} \bibinfo{year}{2020}\natexlab{}.
\newblock \showarticletitle{Optimization of IMU sensor placement for the measurement of lower limb joint kinematics}.
\newblock \bibinfo{journal}{\emph{Sensors}} \bibinfo{volume}{20}, \bibinfo{number}{21} (\bibinfo{year}{2020}), \bibinfo{pages}{5993}.
\newblock


\bibitem[Oishi et~al\mbox{.}(2025)]%
        {oishi2025wimusim}
\bibfield{author}{\bibinfo{person}{Nobuyuki Oishi}, \bibinfo{person}{Phil Birch}, \bibinfo{person}{Daniel Roggen}, {and} \bibinfo{person}{Paula Lago}.} \bibinfo{year}{2025}\natexlab{}.
\newblock \showarticletitle{WIMUSim: simulating realistic variabilities in wearable IMUs for human activity recognition}.
\newblock \bibinfo{journal}{\emph{Frontiers in Computer Science}}  \bibinfo{volume}{7} (\bibinfo{year}{2025}), \bibinfo{pages}{1514933}.
\newblock


\bibitem[Pannurat et~al\mbox{.}(2017)]%
        {pannurat2017analysis}
\bibfield{author}{\bibinfo{person}{Natthapon Pannurat}, \bibinfo{person}{Surapa Thiemjarus}, \bibinfo{person}{Ekawit Nantajeewarawat}, {and} \bibinfo{person}{Isara Anantavrasilp}.} \bibinfo{year}{2017}\natexlab{}.
\newblock \showarticletitle{Analysis of optimal sensor positions for activity classification and application on a different data collection scenario}.
\newblock \bibinfo{journal}{\emph{Sensors}} \bibinfo{volume}{17}, \bibinfo{number}{4} (\bibinfo{year}{2017}), \bibinfo{pages}{774}.
\newblock


\bibitem[Ray et~al\mbox{.}(2024a)]%
        {ray2024har}
\bibfield{author}{\bibinfo{person}{Lala Shakti~Swarup Ray}, \bibinfo{person}{Daniel Gei{\ss}ler}, \bibinfo{person}{Mengxi Liu}, \bibinfo{person}{Bo Zhou}, \bibinfo{person}{Sungho Suh}, {and} \bibinfo{person}{Paul Lukowicz}.} \bibinfo{year}{2024}\natexlab{a}.
\newblock \showarticletitle{ALS-HAR: Harnessing Wearable Ambient Light Sensors to Enhance IMU-Based Human Activity Recognition}. In \bibinfo{booktitle}{\emph{International Conference on Pattern Recognition}}. Springer, \bibinfo{pages}{133--147}.
\newblock


\bibitem[Ray et~al\mbox{.}(2023)]%
        {ray2023selecting}
\bibfield{author}{\bibinfo{person}{Lala Shakti~Swarup Ray}, \bibinfo{person}{Bo Zhou}, \bibinfo{person}{Sungho Suh}, {and} \bibinfo{person}{Paul Lukowicz}.} \bibinfo{year}{2023}\natexlab{}.
\newblock \showarticletitle{Selecting the motion ground truth for loose-fitting wearables: Benchmarking optical mocap methods}. In \bibinfo{booktitle}{\emph{Proceedings of the 2023 ACM International Symposium on Wearable Computers}}. \bibinfo{pages}{27--32}.
\newblock


\bibitem[Ray et~al\mbox{.}(2024b)]%
        {ray2024comprehensive}
\bibfield{author}{\bibinfo{person}{Lala Shakti~Swarup Ray}, \bibinfo{person}{Bo Zhou}, \bibinfo{person}{Sungho Suh}, {and} \bibinfo{person}{Paul Lukowicz}.} \bibinfo{year}{2024}\natexlab{b}.
\newblock \showarticletitle{A comprehensive evaluation of marker-based, markerless methods for loose garment scenarios in varying camera configurations}.
\newblock \bibinfo{journal}{\emph{Frontiers in Computer Science}}  \bibinfo{volume}{6} (\bibinfo{year}{2024}), \bibinfo{pages}{1379925}.
\newblock


\bibitem[Sara et~al\mbox{.}(2023)]%
        {sara2023effect}
\bibfield{author}{\bibinfo{person}{Lauren~K Sara}, \bibinfo{person}{Jereme Outerleys}, {and} \bibinfo{person}{Caleb~D Johnson}.} \bibinfo{year}{2023}\natexlab{}.
\newblock \showarticletitle{The effect of sensor placement on measured distal tibial accelerations during running}.
\newblock \bibinfo{journal}{\emph{Journal of Applied Biomechanics}} \bibinfo{volume}{39}, \bibinfo{number}{3} (\bibinfo{year}{2023}), \bibinfo{pages}{199--203}.
\newblock


\bibitem[Str{\"o}mb{\"a}ck et~al\mbox{.}(2020)]%
        {stromback2020mm}
\bibfield{author}{\bibinfo{person}{David Str{\"o}mb{\"a}ck}, \bibinfo{person}{Sangxia Huang}, {and} \bibinfo{person}{Valentin Radu}.} \bibinfo{year}{2020}\natexlab{}.
\newblock \showarticletitle{Mm-fit: Multimodal deep learning for automatic exercise logging across sensing devices}.
\newblock \bibinfo{journal}{\emph{Proceedings of the ACM on Interactive, Mobile, Wearable and Ubiquitous Technologies}} \bibinfo{volume}{4}, \bibinfo{number}{4} (\bibinfo{year}{2020}), \bibinfo{pages}{1--22}.
\newblock


\bibitem[Tan et~al\mbox{.}(2019)]%
        {tan2019influence}
\bibfield{author}{\bibinfo{person}{Tian Tan}, \bibinfo{person}{David~P Chiasson}, \bibinfo{person}{Hai Hu}, {and} \bibinfo{person}{Peter~B Shull}.} \bibinfo{year}{2019}\natexlab{}.
\newblock \showarticletitle{Influence of IMU position and orientation placement errors on ground reaction force estimation}.
\newblock \bibinfo{journal}{\emph{Journal of biomechanics}}  \bibinfo{volume}{97} (\bibinfo{year}{2019}), \bibinfo{pages}{109416}.
\newblock


\bibitem[Tsukamoto et~al\mbox{.}(2023)]%
        {tsukamoto2023best}
\bibfield{author}{\bibinfo{person}{Akihisa Tsukamoto}, \bibinfo{person}{Naoto Yoshida}, \bibinfo{person}{Tomoko Yonezawa}, \bibinfo{person}{Kenji Mase}, {and} \bibinfo{person}{Yu Enokibori}.} \bibinfo{year}{2023}\natexlab{}.
\newblock \showarticletitle{Where Are the Best Positions of IMU Sensors for HAR?-Approach by a Garment Device with Fine-Grained Grid IMUs}. In \bibinfo{booktitle}{\emph{Adjunct Proceedings of the 2023 ACM International Joint Conference on Pervasive and Ubiquitous Computing \& the 2023 ACM International Symposium on Wearable Computing}}. \bibinfo{pages}{445--450}.
\newblock


\bibitem[Uhlenberg and Amft(2024)]%
        {uhlenberg2024mount}
\bibfield{author}{\bibinfo{person}{Lena Uhlenberg} {and} \bibinfo{person}{Oliver Amft}.} \bibinfo{year}{2024}\natexlab{}.
\newblock \showarticletitle{Where to mount the IMU? Validation of joint angle kinematics and sensor selection for activities of daily living}.
\newblock \bibinfo{journal}{\emph{Frontiers in Computer Science}}  \bibinfo{volume}{6} (\bibinfo{year}{2024}), \bibinfo{pages}{1347424}.
\newblock


\bibitem[Young et~al\mbox{.}(2011)]%
        {young2011imusim}
\bibfield{author}{\bibinfo{person}{Alexander~D Young}, \bibinfo{person}{Martin~J Ling}, {and} \bibinfo{person}{Damal~K Arvind}.} \bibinfo{year}{2011}\natexlab{}.
\newblock \showarticletitle{IMUSim: A simulation environment for inertial sensing algorithm design and evaluation}. In \bibinfo{booktitle}{\emph{Proceedings of the 10th ACM/IEEE International Conference on Information Processing in Sensor Networks}}. IEEE, \bibinfo{pages}{199--210}.
\newblock


\bibitem[Zolfaghari et~al\mbox{.}(2024)]%
        {zolfaghari2024sensor}
\bibfield{author}{\bibinfo{person}{Parham Zolfaghari}, \bibinfo{person}{Vitor~Fortes Rey}, \bibinfo{person}{Lala Ray}, \bibinfo{person}{Hyun Kim}, \bibinfo{person}{Sungho Suh}, {and} \bibinfo{person}{Paul Lukowicz}.} \bibinfo{year}{2024}\natexlab{}.
\newblock \showarticletitle{Sensor data augmentation from skeleton pose sequences for improving human activity recognition}. In \bibinfo{booktitle}{\emph{2024 International Conference on Activity and Behavior Computing (ABC)}}. IEEE, \bibinfo{pages}{1--8}.
\newblock


\bibitem[Zuo et~al\mbox{.}(2024)]%
        {zuo2024loose}
\bibfield{author}{\bibinfo{person}{Chengxu Zuo}, \bibinfo{person}{Yiming Wang}, \bibinfo{person}{Lishuang Zhan}, \bibinfo{person}{Shihui Guo}, \bibinfo{person}{Xinyu Yi}, \bibinfo{person}{Feng Xu}, {and} \bibinfo{person}{Yipeng Qin}.} \bibinfo{year}{2024}\natexlab{}.
\newblock \showarticletitle{Loose inertial poser: Motion capture with IMU-attached loose-wear jacket}. In \bibinfo{booktitle}{\emph{Proceedings of the IEEE/CVF Conference on Computer Vision and Pattern Recognition}}. \bibinfo{pages}{2209--2219}.
\newblock


\end{thebibliography}


\end{document}